\documentclass[12pt]{article}
\pdfoutput=1
\usepackage{latexsym}\usepackage{epsfig,amssymb,euscript}
\usepackage{amsmath} \usepackage{bbm,slashed}
\usepackage[usenames,dvipsnames]{color}
\usepackage{cite}
\usepackage{graphicx}
\usepackage{hyperref}

\usepackage[hang,small,bf]{caption}
\newcommand{\nn} {\nonumber\\}

\def\sfU{\mathsf{U}}

\def\ImO{\mathrm{Im}O}
\def\ReO{\mathrm{Re}O}

\begin{document}

\begin{titlepage}

\begin{flushright}
\end{flushright}

\bigskip
\def\thefootnote{\fnsymbol{footnote}}

\begin{center}
\vskip -10pt
{\LARGE
{\bf
Analytic pseudo-Goldstone bosons
}
}
\end{center}

\bigskip
\begin{center}
{\large 
Riccardo Argurio$^{1}$, Andrea Marzolla$^{1}$,\\ \vskip 5pt
Andrea Mezzalira$^{2,3}$ and Daniele Musso$^4$}

\end{center}

\renewcommand{\thefootnote}{\arabic{footnote}}

\begin{center}
\vspace{0.2cm}
$^1$ {Physique Th\'eorique et Math\'ematique and International Solvay Institutes \\ Universit\'e Libre de Bruxelles, C.P. 231, 1050 Brussels, Belgium\\}
$^2$ {Institute Lorentz for Theoretical Physics, Leiden University\\P.O. Box 9506, Leiden 2300RA, The Netherlands\\}
$^3$ {School of Mathematics, Trinity College, Dublin 2, Ireland\\}
$^4$ {Abdus Salam International Centre for Theoretical Physics (ICTP)\\
Strada Costiera 11, I 34151 Trieste, Italy}

\vskip 5pt

\end{center}

\noindent
\begin{center} {\bf Abstract} \end{center}
\noindent
We consider the interplay between explicit and spontaneous symmetry breaking in strongly coupled field theories. Some well-known statements, such as the Gell-Mann-Oakes-Renner relation, descend directly from the Ward
identities and have thus a general relevance. Such Ward identities are  recovered in gauge/gravity dual setups through holographic renormalization. In a simple paradigmatic three dimensional toy-model, we find analytic expressions for the two-point correlators  which match all the  quantum field theoretical expectations. Moreover, we have access to the full spectrum, which is reminiscent of linear confinement.

\vspace{1.6 cm}
\vfill

\end{titlepage}

\setcounter{footnote}{0}

\tableofcontents

\section{Introduction}

Gaining systematic understanding of symmetry breaking at strong coupling 
constitutes one of the main open fronts in contemporary Quantum Field Theory investigations.
Conceptual questions are side by side to diverse applications in motivating this field of research. Supersymmetry breaking
in strongly coupled hidden sectors and high-$T_c$ superconductors are just two important instances of such a wide range of applications. 

The purpose of the paper is to delve into the theoretical aspects of symmetry breaking at strong coupling 
to provide neat and general descriptions of the dynamics 
in generic cases featuring concomitant explicit and spontaneous breaking.
The key theoretical tools, not surprisingly, are the correlators of the gauge invariant 
quantities and the relations among them descending 
from the original symmetry and its breaking, namely the Ward identities. 

We first rely on a purely field theoretical 
setting illustrating how the Ward identity structure, supplemented with appropriate and generically valid 
consistency requirements, can alone provide a description of correlators 
whose low energy and momentum behavior is characterized by pseudo-Goldstone bosons.
As the Goldstone theorem guarantees the presence of Goldstone modes whenever the breaking 
of a global internal symmetry is purely spontaneous \cite{Goldstone:1962es},
similarly the QFT framework is capable of grasping precisely the essential dynamics of pseudo-Goldstone modes whenever 
the breaking is  both explicit and spontaneous; specifically, when the explicit component is parametrically small \cite{GellMann:1968rz,Gasser:1982ap}. 

The standard theoretical difficulties of strong coupling manifest themselves in the contest of 
symmetry breaking since both the fields responsible for and arising from the symmetry breaking dynamics 
are in general composite and impossible to treat in a perturbative fashion. 
A fruitful way of approaching the problem is in gauge invariant, ``operatorial'' terms,
where one describes the structure of correlators and expectation values of observables. 

The above setting is particularly suitable to be treated in a gauge/gravity context  \cite{Maldacena:1997re,Witten:1998qj,Gubser:1998bc}.
The holographic renormalization of a gauge/gravity model precisely implements 
the Ward identities satisfied by correlators of the dual strongly coupled field theory \cite{Bianchi:2001de,Bianchi:2001kw}. 
In this framework we describe one of the simplest holographic models for a prototypical 
$U(1)$ symmetry breaking allowing for a complete analytic treatment. 
In this model we are able to determine exactly and analytically the correlators of the operators 
in the symmetry breaking sector. From them we can extract the complete spectrum of composite bound states. 
Of particular interest is the lightest mode, i.e.~the pseudo-Goldstone boson. 
The relevant input scales of the problem are actually the parameters controlling the explicit and spontaneous 
components of the symmetry breaking, respectively. The output data is represented by the quantities characterizing 
the spectrum, namely the masses and the residues of the poles. For the pseudo-Goldstone bosons, 
these quantities organize according to the 
celebrated Gell-Mann-Oakes-Renner relation \cite{GellMann:1968rz} in the limit of small explicit breaking. 
The toy-model has the additional interesting feature of having, in the purely spontaneous case, a spectrum reproducing linear confinement, i.e.~massive bound states equally spaced in the squared masses. 

Though the analytic power of the approach is somehow restricted to our specific toy-model, we believe that the qualitative picture generalizes to a full class of 
holographic bottom-up models encompassing different space-time dimensions and 
diverse operator content. In particular, the physical understanding emerging from the present analysis
makes contact with numerous previous studies in the holographic literature, 
both in the top-down \cite{Babington:2003vm,Kruczenski:2003uq,Evans:2004ia,Casero:2007ae,Bergman:2007pm} 
and in the bottom-up spirit \cite{Erlich:2005qh,DaRold:2005zs,Karch:2006pv,Erlich:2008gp}.

The paper is organized as follows. In section 2 we derive purely in field theory and in full generality some properties 
that the correlators have to satisfy as a consequence of the Ward identities in the presence of symmetry breaking. 
One such property is the Gell-Mann-Oakes-Renner (GMOR) relation for the pseudo-Goldstone boson (PGB) in the limit of small explicit breaking. 
In section 3 we present our three-dimensional toy-model and show how holographic renormalization reproduces exactly the Ward identities. 
We stress that only near-boundary analysis is utilized to reach these conclusions. 
In section 4 we solve the bulk equations for the fluctuations and find exact analytical expressions for the correlators 
for any values of the explicit and spontaneous symmetry breaking parameters. 
We then explore several limits, extracting the spectrum, and showing that the correlators satisfy non-trivially all the consistency checks.  
In the two appendices we present respectively the generalization of the holographic model to arbitrary dimensions, and the departure from the GMOR relation when the explicit breaking is no longer small.

\section{Pseudo-Goldstone bosons and GMOR relations from Ward identities}

Consider a field theory which is invariant under a symmetry, which for simplicity we take to be a $U(1)$. 
It has a conserved current $J^\mu$. We can add to the action a term which explicitly breaks the symmetry
\begin{equation}
S_\mathrm{tot} = S_\mathrm{inv} + \int d^d x\  \frac12\, m\, O_\phi\, +\, c.c.\ ,
\end{equation}
where $O_\phi$ is a scalar operator of dimension $\Delta$  which is charged under the $U(1)$. 
Again for simplicity, we take its charge to be unity and the explicit breaking parameter $m$ to be real. The current is no longer conserved, rather one has the operator identity
$\partial_\mu J^\mu =  m\, \ImO_\phi$. Furthermore, if in the general case the operator  develops a VEV $\langle O_\phi\rangle=v$, which we take to be real, the Ward identity
\begin{equation}
\langle \partial_\mu J^\mu(x) \ImO_\phi(0)\rangle = m \langle \ImO_\phi(x) \ImO_\phi(0)\rangle+i \langle \ReO_\phi\rangle \delta^d(x)
\end{equation}
implies that the following two-point functions all depend on a single non-trivial function $f(\Box)$, namely
%
\begin{align}
\langle \ImO_\phi \ImO_\phi\rangle &=- i f(\Box)\ ,\label{oocorr} \\
\langle \partial_\mu J^\mu \ImO_\phi\rangle & = -i m f(\Box) + iv\ ,\label{jocorr} \\
\langle \partial_\mu J^\mu \partial_\nu J^\nu\rangle & =- im^2 f(\Box) +imv\ ,\label{jjcorr}
\end{align}
where we have kept the delta function implicit. The last correlator is just a consequence of the operator identity.

When $v=0$ (pure explicit breaking case), then also the second correlator is a trivial consequence of the operator identity. 
On the other hand, when $m=0$ (pure spontaneous breaking case), the second correlator is a constant directly determined by the Ward identity. 
It implies the presence of a massless pole in $\langle J_\mu \ImO_\phi\rangle$. 
The same massless excitation has however to appear in \eqref{oocorr}, though  its description requires an analysis of the (IR) dynamics.

When both $v\neq 0$ and $m\neq 0$, we see that \eqref{jocorr} has both features, a term related to \eqref{oocorr} and a constant term.
On the other hand, since the symmetry is broken explicitly, we do not expect a massless mode in the spectrum contributing to this set of correlators. 
As we will see, requiring continuity in the $m\to 0$ limit allows us to find the Gell-Mann-Oakes-Renner relation for the mass of the pseudo-Goldstone boson \cite{GellMann:1968rz}.%
\footnote{Standard derivations of such a relation can be found, e.g., in \cite{Gasser:1982ap,Giusti:1998wy}.
Though close in spirit, the derivation we present in this section is original, to our knowledge. 
Its starting point is precisely the outcome of the holographic analysis of the next section.  See also, e.g., \cite{Evans:2004ia} for the equally standard derivation based on the effective action.}

In momentum space we write%
\footnote{Even though $\Box$ corresponds to $-k^2$, with a venial abuse of language, we keep denoting the function $f$ with the same symbol also in momentum space.}
\begin{equation}
\langle \ImO_\phi \ImO_\phi\rangle = -if(k^2)\ , \qquad 
ik_\mu \langle  J^\mu \ImO_\phi\rangle = -im f(k^2) +iv\ .
\end{equation}
Note that  the dimensions are  $[f]=2\Delta-d$, $[v]=\Delta$ and $[m]=d-\Delta$.

Using Lorentz invariance of the vacuum, which imposes $\langle  J^\mu \ImO_\phi\rangle = k^\mu g(k^2)$, the second relation leads to
\begin{equation}
\langle  J^\mu \ImO_\phi\rangle =- \frac{k^\mu}{k^2}(m f(k^2) - v)\ .
\end{equation}
We immediately see that there cannot be a massless excitation in the $\langle \ImO_\phi \ImO_\phi\rangle$ channel, 
otherwise there would be a double pole in $\langle  J^\mu \ImO_\phi\rangle$. Moreover, 
the massless pole in the above correlator should be spurious, which means that $f(k^2)$ has to satisfy
\begin{equation}
m f(0) - v=0\ .\label{scheme}
\end{equation}
In general, we could wonder whether there exist local finite counter-terms that modify the constant part of $f(k^2)$ in order 
to impose the above condition through a scheme choice. This is a question that depends on the specifics of the model, in particular $d$ and $\Delta$. We will discuss an example where
there are no gauge-invariant, local, finite counter-terms  and \eqref{scheme} has to emerge directly and unambiguously from the computations.
Note that what we have determined until now is true for any values of $m$ and $v$.  

For $m$ and $k$ both small with respect to $\sqrt{v}$, we can approximate $f$ by a pole corresponding to the PGB of mass $M$
\begin{equation}
f(k^2)\ \simeq \ \frac{\mu}{k^2+M^2} -\frac{\mu}{M^2} +\frac{v}{m}\ ,\label{fpgb}
\end{equation}
where we have implemented the condition \eqref{scheme}, and the residue $\mu$ is a dynamical quantity of dimension $2\Delta-d+2$. 
We now require that in the $m\to 0$ limit, $f(k^2)$ goes over smoothly to what we expect in the pure spontaneously broken case. 
Namely, we expect $\mu$ to be (roughly) constant in the limit, as of course $v$, while $M^2 \to 0$, so that 
\begin{equation}
f(k^2) \to \frac{\mu}{k^2}\ ,
\end{equation}
up to possibly an additive finite constant. From \eqref{fpgb} we see that this is possible only if there is a relation between all the constants such that 
\begin{equation}
M^2 = \frac{\mu}{v}\, m \ .\label{gmor}
\end{equation}
This is the generalization of the GMOR relation \cite{GellMann:1968rz}, 
which indeed states that the squared mass of the PGB scales linearly with the small parameter which breaks explicitly
the symmetry. The two other constants entering the expression are both of the order 
of the dynamical scale generating the VEV, i.e. the spontaneous breaking of the symmetry.

Note that since $\mu$ has to be positive because of unitarity, then the signs (and more generally the phases) of $m$ and $v$ 
have to be correlated in order to avoid tachyonic PGBs. This can be understood by the 
fact that the small explicit breaking removes the degeneracy of the vacua, and thus the phase 
of the VEV $v$ is no longer arbitrary but has to be aligned with the true vacuum selected by $m$.

A last remark is that the usual way in which the GMOR relation is stated is in terms of the 
residue of the $\langle J^\mu J^\nu\rangle$ correlator
\begin{equation}
\langle J^\mu J^\nu\rangle =- \frac{i\mu_J}{k^2+M^2}k^\mu k^\nu +\dots\ ,
\end{equation}
where $\mu_J$ is related to the square of the ``PGB decay constant".

Note that implementing \eqref{gmor} in the correlators we get 
\begin{equation}
f(k^2)=\frac{\mu}{k^2+M^2}\ , \qquad\qquad mf(k^2)-v=-\frac{vk^2}{k^2+M^2}\ ,
\end{equation}
so that, at $k^2=-M^2$, we have
\begin{equation}
\mu_J M^4 = mv M^2\ ,
\end{equation}
which leads to 
\begin{equation}
M^2 = m\frac{v}{\mu_J}\ ,
\end{equation}
namely the usual GMOR relation, which is thus completely equivalent to \eqref{gmor}.

Above we have kept both $d$ and $\Delta$ arbitrary, and the relation is  valid in all generality. 
In the following we will discuss a specific model where $d=3$ and $\Delta=2$, so that $m$ has 
indeed the dimension of a mass.

\section{Holographic toy-model}

To illustrate the interplay between explicit and spontaneous symmetry
breaking, we use as a toy-model a simplified, Abelian version of the
model used in \cite{Amado:2013xya,Argurio:2015via}, which coincides with the model of the
very first holographic superconductor \cite{Gubser:2008px,Hartnoll:2008vx}.
The Ward identity structure emerges through the precise holographic renormalization  procedure \cite{Bianchi:2001de,Bianchi:2001kw} which therefore constitutes our first task.

We start considering the action
\begin{equation}
S=\int d^4x \sqrt{-g}\left\{ -\frac14 F^{MN}F_{MN}-D_M\phi^* D^M\phi +2 \phi^*\phi \right\}\ ,
\label{sbulk}
\end{equation}
where $D_M \phi=\partial_M \phi -i A_M \phi$, the metric is $AdS$ and we choose the most general profile for $\phi$, namely
\begin{equation}\label{bg}
ds^2= \frac{1}{z^2} (dz^2+dx_\mu d x^\mu)\ , \qquad \qquad
\phi_b= mz+vz^2\ .
\end{equation}
We keep Lorentz invariance unbroken, and hence we have a vanishing background for $A_\mu$.
Moreover we have chosen the squared mass of the scalar to be $-2$ corresponding to a dual operator
having dimension $\Delta = 2$; the physics we describe is generic and not specifically related 
to such a choice which has been made (as commented later) for technical convenience. Note also 
that we are inside the window where an alternative quantization could be considered \cite{Klebanov:1999tb}.

We can now compute the action for the fluctuations above the background \eqref{bg}, i.e.~we linearize 
$\phi=\frac{1}{\sqrt2}(\phi_b +\rho + i \pi)$, where $\phi_b$ is assumed to be real for simplicity (and, as we have already remarked, also for consistency). 
The rescaling prefactor $\sqrt{2}$ with respect to the generic shape of the scalar profile given in \eqref{bg} is designed to match the results with the previous section.
As shown in \cite{Argurio:2015via}, independently of the specific form of the background, in the $A_z=0$ gauge 
the regularized action up to quadratic order can be written as the following boundary term
\begin{equation}
S_\mathrm{reg}=-\int d^3 x \left\{ -\frac{1}{z^2}\partial_z\phi_b\rho-\frac12 A_\mu \partial_z A_\mu -\frac{1}{2z^2}\left(\rho\partial_z\rho+\pi\partial_z\pi \right)\right\}\ .\label{sreg}
\end{equation}
It is possible to rewrite the above expression using the equation of motion coming from the variation 
of \eqref{sbulk} with respect to $A_z$, which in the ``radial'' $A_z=0$ gauge reads
\begin{equation}
-z^4 \partial_z\partial_\mu A_\mu+iz^2 \phi \partial_z \phi^*-iz^2 \phi^* \partial_z \phi=0\ .
\end{equation}
Linearizing and taking $A_\mu=A_\mu^t+\partial_\mu A_l$ (namely splitting the longitudinal and transverse part),  it rewrites
\begin{equation}
z^2\partial_z\Box A_l - \phi_b\partial_z\pi+\partial_z\phi_b\pi=0\ .
\end{equation}
Noting that the second term of \eqref{sreg} has a longitudinal part that can be rewritten, 
after partial integration, as $\frac12 A_l \partial_z \Box A_l$, the regularized action for the longitudinal part and the scalars becomes
\begin{equation}
S_\mathrm{reg}=-\int d^3 x \left\{ -\frac{1}{z^2}\partial_z\phi_b\rho-
\frac{1}{2z^2} A_l \left(\partial_z\phi_b\pi-\phi_b\partial_z\pi\right)-\frac{1}{2z^2}\left(\rho\partial_z\rho+\pi\partial_z\pi \right)\right\}\ .\label{sreg2}
\end{equation}
We now expand the fluctuations as
\begin{equation}\label{nbe}
A_l=A_0+A_1 z +\dots\ , \qquad  \rho = \rho_0 z +\rho_1 z^2+\dots\ , \qquad \pi=\pi_0 z + \pi_1 z^2+\dots\ .
\end{equation}
Eq.~\eqref{sreg2} then becomes
\begin{align}
S_\mathrm{reg}=-\int d^3 x& \left\{-\frac{1}{z}m\rho_0-2v\rho_0 -m\rho_1\right.\label{sreg3}\\
&\quad\left.-\frac{1}{2z} (\rho_0^2+\pi_0^2)-\frac32 (\rho_0\rho_1+\pi_0\pi_1)+ \frac12A_0 (m\pi_1-v \pi_0)
\right\}\ .\nonumber
\end{align}
The counter-term needed to cancel the divergencies is
\begin{equation}
S_\mathrm{ct}=-\int d^3x \sqrt{-\gamma}\phi\phi^* = -\int d^3x \left\{ \frac{1}{z^3} \phi_b\rho+ \frac{1}{2z^3} (\rho^2+\pi^2)\right\}  \ .\label{sct}
\end{equation}
Note that we  neglect the constant term as it would only be relevant with dynamical gravity.
After adding the counter-term \eqref{sct} to \eqref{sreg3}, we obtain the renormalized action
\begin{equation}
S_\mathrm{ren}=-\int d^3 x \left\{-v\rho_0 -\frac12\rho_1\rho_0
-\frac12\pi_1(\pi_0-mA_0) -\frac12v A_0\pi_0\right\}\ .\label{sren0}
\end{equation}

Let us now discuss gauge invariance. 
Under a gauge transformation that preserves $A_z=0$ we have
\begin{equation}
\delta A_l=\alpha, \qquad \delta \phi=i\alpha \phi\ .
\end{equation}
The first transformation above tells that $\alpha$ should be considered of the same order as the 
fluctuations $A_l$ and $\rho, \pi$. It then implies that the gauge variations of $\rho,\pi$ have actually terms of first and second order 
\begin{equation}
\delta\rho = -\alpha\pi \ , \qquad \delta\pi =  \alpha\phi_b + \alpha \rho\ .
\end{equation}
On the coefficients of the near-boundary expansions \eqref{nbe}, the transformations read
\begin{equation}
\delta A_0 = \alpha\ ,\qquad \delta A_1=0\ ,\qquad \delta \rho_0 = -\alpha\pi_0\ , \qquad
\delta \rho_1 = -\alpha\pi_1\ , \nonumber
\end{equation}
\begin{equation}
\delta \pi_0 = \alpha m + \alpha \rho_0\ , \qquad
\delta \pi_1 = \alpha v + \alpha \rho_1\ .
\end{equation}
With the transformations given above, one can check that all the actions
$S_\mathrm{reg}$, $S_\mathrm{ct}$ and $S_\mathrm{ren}$ are gauge invariant. 
We note that gauge invariance requires the cancellation between the variations of 
the linear and quadratic parts of the actions, and we have of course neglected 
orders higher than quadratic%
\footnote{It is possible also to parametrize the complex scalar in terms of its modulus and phase as in \cite{Argurio:2014rja};
the latter parametrization, being well-adapted to gauge transformations (which consist in shifts of the phase), features manifest gauge invariance
without mixing among different orders in the fluctuations.}
(i.e. in the variations of the quadratic part of 
the action, only the terms of first order in the variations of $\rho,\pi$ are considered).

In the renormalized action \eqref{sren0} it is manifest that the $\rho$ sector decouples from the $A_l,\pi$ sector. We will not be concerned with the former, except for the obvious fact that the linear, $\rho$-dependent term gives the VEV of the operator.

In order to solve for $\pi_1$ in terms of the sources $\pi_0$ and $A_0$, 
one should be careful that the deep bulk (IR) boundary conditions will impose relations between gauge invariant quantities. At linear order, 
the gauge invariant combinations are $\pi_0-mA_0$ and $\pi_1-vA_0$. 
As a consequence, one can express the subleading mode of $\pi$ in terms of the sources as
\begin{equation}
\pi_1= v A_0 +f(\Box)(\pi_0-m A_0)\ .	\label{nonlocf}
\end{equation}
The  renormalized action for this sector can be rewritten accordingly
\begin{equation}
S_\mathrm{ren}=-\int d^3 x \;\Big[\, -\frac12(\pi_0-mA_0)f(\Box)(\pi_0-mA_0)
-vA_0\pi_0+\frac12mv A_0A_0 \,\Big] 	\ .\label{sren}
\end{equation}
We observe that we have a term linear in $m$, which encodes the operator 
identities that are present when the symmetry is explicitly broken. Then we have a term linear 
in $v$, which embodies the Ward identities when the symmetry is spontaneously broken. 
Eventually we have a term which is linear both in $m$ and $v$ and is necessary in order to recover the proper Ward identities in the case of concomitant spontaneous and explicit breaking.

We now derive the holographic correlators, assuming that the terms coupling the sources to the operators are 
\begin{equation}
\int d^3x \left(\rho_0 \ReO_\phi + \pi_0 \ImO_\phi -A_0 \partial_\mu J^\mu\right)\ ,
\end{equation}
where the last term comes from integration by parts. 
We also assume the usual holographic prescription in its Wick-rotated, Lorentzian version.
For instance for the VEV of $\ReO_\phi$, we have
\begin{equation}
\langle \ReO_\phi\rangle = \frac{\delta iS_\mathrm{ren}}{\delta i\rho_0} = v\ .
\end{equation}
For the two-point correlators in the longitudinal sector, we have:
\begin{align}
\langle \ImO_\phi\ImO_\phi\rangle &=  \frac{\delta^2 iS_\mathrm{ren}}{\delta i\pi_0\delta i\pi_0}=-if(\Box)\ ,\label{corrpi0pi0}\\
\langle\partial_\mu J^\mu\ImO_\phi\rangle &=  -\frac{\delta^2 iS_\mathrm{ren}}{\delta iA_0\delta i\pi_0}=-imf(\Box)+iv\ , \\
\langle\partial_\mu J^\mu\partial_\nu J^\nu\rangle &=  \frac{\delta^2 iS_\mathrm{ren}}{\delta iA_0\delta iA_0}=-im^2f(\Box)+imv\  .
\end{align}
These are exactly the equations \eqref{oocorr}--\eqref{jjcorr} obtained from QFT arguments,  that we used to derive the GMOR relations.
Now we proceed to compute holographically the non-trivial function $f(k^2)$ and show that it reproduces the physics that one expects on general grounds.

\section{Analytical study of the fluctuations and 2-point correlators}

In this section we study  the bulk equations of motion for the fluctuations, in order to extract exact expressions for the correlators. 
We will thus be able to verify explicitly that they satisfy the non-trivial conditions discussed in section 2. 
 
We start again from the action \eqref{sbulk}. Allowing for the moment for a possible backreaction of the scalar's background profile on the metric,
the latter is now defined by
\begin{align}
d s^2 = \; &
g_{M\!N}\, dx^M dx^N = 
	\frac{1}{z^2} \left(dz^2+ F(z)	\eta_{\mu\nu} dx^\mu dx^\nu 	\right) \ ,		\label{AdS0}
\end{align}
where the warp factor $F(z)$ is such that $F(0)=1$ (i.e. asymptotically $AdS$)
while it decreases monotonically for increasing $z$. 
We assume that $\phi$ has a profile along $z$, that produces a non-trivial $F$ through backreaction.
 In this way
we can express the fields in terms of fluctuations over the background
\begin{align}
A_M(z,\,x) dx^M & = A_z (z,\,x) dz + A_\mu (z,\,x) dx^\mu
\ ,
\\
\phi(z,\,x) & = \frac{1}{\sqrt{2}}\left(  \phi_b (z) + \varphi (z,\,x)\right)  
\ .
\label{back0}
\end{align}
We furthermore gauge-fix $A_z=0$, and we eventually derive the following equations of motion for the fluctuations
\begin{align}
& 	\frac{z^2}{F^{1/2}}\partial_z(F^{1/2}\partial_z A_\mu) 	+\frac{z^2}{F}(\Box A_\mu 	-\partial_{\mu} \partial_\nu A^\nu)	-\frac i2 \, \phi_b\, \partial_\mu ( \varphi - \varphi^* )  -\phi_b^2\, A_\mu 	=0 \ , \phantom{\Big]} \nn 
&	\frac{z^2}{F}\partial_z \partial_\mu A^\mu 	+\frac i2 \, \phi_b\,	 \partial_z ( \varphi - \varphi^* ) 	-\frac i2 \partial_z \phi_b( \varphi - \varphi^* ) 	=0 \ , 	\phantom{\Big]} \label{EoM0} \\ 
&	\frac{z^4}{F^{3/2} }\partial_z\left(\frac{F^{3/2}}{z^2}\partial_z \varphi\right) 		+\frac{z^2}{F} \Box \varphi 	-i \frac{z^2}{F} \phi_b \partial^\mu A_\mu 	+2\varphi 	=0 \ .	\phantom{\Big]} \nonumber
\end{align}
We then split the gauge field in its transverse and longitudinal parts as follows
\begin{equation}
A_\mu = A_\mu^t + \partial_\mu A_l \ , 		\qquad\qquad 	\partial^\mu A_\mu^t = 0 \ ,	\label{splitTL}
\end{equation}
and we define $\varphi = \rho+i \pi $, so that eqs.~(\ref{EoM0}) split into five equations: two equations for $A_\mu^t$ and $\rho$, respectively, decoupled from each other and from the rest, which we will not consider further; and  a set of three coupled equations for $A_l$ and $\pi$, namely
\begin{align}
&	\frac{z^2}{F^{1/2}}\partial_z(F^{1/2}\partial_z A_l) +\phi_b \pi 	-\phi_b^2 A_l	=0 \ ,	\phantom{\Big]} \label{eom-a} \\ 
& 	\frac{z^2}{F} \Box \partial_z A_l 	-\phi_b  \partial_z \pi 	+\partial_z\phi_b \pi 	=0 \ , \phantom{\Big]} \label{eom-cstr} \\ 
& 	\frac{z^4}{F^{3/2} }\partial_z\left(\frac{F^{3/2}}{z^2}\partial_z \pi\right)	+2 \pi 	+\frac{z^2}{F} \Box \pi 	- \frac{z^2}{F} \phi_b \Box A_l 	=0\ .	\phantom{\Big]} \label{eom-pi}
\end{align}
We can extract $\pi$ from the first equation
\begin{equation}
\pi = \phi_b A_l - 	\frac{z^2}{F^{1/2}\phi_b}\partial_z(F^{1/2}\partial_z A_l)\ .
\label{pisol}
\end{equation}
Note that gauge transformations, at linear order, are given by $\delta A_l=\alpha$ and $\delta \pi=\phi_b \alpha$, where $\alpha$ does not depend on $z$ because of the gauge fixing condition $A_z=0$. 
We then see that both $A_l'$ and $\pi-\phi_b A_l$ are gauge invariant quantities.

We then plug \eqref{pisol} into \eqref{eom-cstr}, and we find 
a second order differential equation for $F^{1/2}A_l'\equiv B$ (as expected from gauge invariance)
\begin{equation}
z^2 B'' +2z B'-\frac{1}{2}z^2 \frac{F'}{F}B'-2z^2\frac{\phi_b'}{\phi_b}B' +\frac{ z^2}{F} \Box B -\phi_b^2B =0 \ .		\label{ODEgen}
\end{equation}
The system of three equations is therefore reduced to a single second order ODE.

We have included backreaction to show that it does not change substantially the equations for the fluctuations. 
Its effects are subdominant, as we will argue below. Hence, let us  consider from now on the case without backreaction, 
i.e.~$F=1$. In this case the scalar profile is $\phi_b=mz+vz^2$, where the leading term encodes  
explicit symmetry breaking, whereas the sub-leading one corresponds to spontaneous symmetry breaking. Eventually equation \eqref{ODEgen} simplifies to 
\begin{equation}
B''-\frac{2v}{m+vz}\,B'-k^2 B-(m+vz)^2 B =0\ , \label{ODEnb}
\end{equation}
and, by the simple change of variable $y=z+\frac mv$, we obtain 
\begin{equation}
B''-\frac{2}{y}B'-k^2 B -v^2 y^2 B =0\ , \label{ODEy}
\end{equation}
which can be recast as a general confluent hypergeometric equation. Its solutions are given in terms of the Tricomi's confluent hypergeometric function $\sfU[a,b;x]$ and the generalized Laguerre polynomial $\mathsf{L}[a,b;x]$
\begin{equation}
B(y) = \exp\!\left[-\frac{v y^2}{2}\right] 
	\left( C_1\; \sfU\Big[ \frac{k^2-v}{4v},\, -\frac 12;\, v y^{2} \Big] +C_2\; \mathsf{L}\Big[\frac{v-k^2}{4v},\, -\frac 32;\, v y^2 \Big] \right) \ .
\end{equation}
In the deep bulk ($y\!\rightarrow\!\infty$), we have $e^{-\frac{v y^2}{2}}\sfU\!\sim\! e^{-\frac{v y^2}{2}}$ whereas $e^{-\frac{v y^2}{2}}\mathsf{L}\!\sim\! e^{+\frac{v y^2}{2}}$. 
Since $\partial_{z} A_l$ is gauge-invariant, we are allowed to impose IR boundary conditions on it, and we choose  bulk normalizability of the solution setting $C_{2}\equiv0$. 
We thus obtain
\begin{equation}
B(y) =  C_1\: e^{-\frac{v}{2}y^2}\; \sfU\Big[\frac{k^2-v}{4v},\,-\frac 12;\, v y^{2} \Big]\ . 	\label{Bdiy}
\end{equation}
Note that this solution has a very fast decrease towards the interior of the bulk, confirming that backreaction will only affect mildly the correlators that we will extract from it.

In this way we have obtained an exact analytical solution for the derivative of the gauge field, but we still have to derive a solution for $\pi$, in order to compute the scalar correlator. 
If we consider the near-boundary expansion for the fluctuations
\begin{align}
\pi & 	= z\,\pi_{0} +z^{2}\,\pi_{1} +\ldots \ , 	\label{piexp}\\
A_l & 	= A_{0} +z\, A_{1} +z^2\,A_2 +z^3A_3 +\ldots \ ,	\label{Alexp}
\end{align}
then we need to know the expressions for $\pi_{0}$ and $\pi_{1}$ in order to compute the scalar correlator. 
Indeed, from \eqref{nonlocf} and \eqref{corrpi0pi0}, we see that
\begin{equation}
\langle \ImO_\phi \ImO_\phi\rangle = -i\frac{\delta \pi_{1}}{\delta\pi_{0}} = -i f(k^2)\ .		\label{2p-corr}
\end{equation}
In other words, the correlator is essentially extracted from \eqref{nonlocf}, that we rewrite here as
\begin{equation}
\pi_{1}-v\, A_{0} = f(k^2) \left(\pi_{0} -mA_0\right) \ .	\label{gaugeinvf}
\end{equation}
From equation~\eqref{pisol} with $F=1$, we can express the gauge invariant combination appearing in eq.~\eqref{gaugeinvf} in terms of $A_l$
\begin{equation}
\pi -\phi_b A_l= -\frac{z^2}{\phi_b}\, A_l''(z) 	\ .
\end{equation}
Order by order near the boundary, through the expansions (\ref{piexp})--(\ref{Alexp}), we obtain
\begin{align}
& \pi_{0} -mA_0 = -\frac{1}{m}\, 2A_2	\ ; \\
& \pi_{1} -v\, A_{0} = m\,A_1 +\frac{v}{m^2}\,2A_2 -\frac{1}{m}\,6A_3 	\ .
\end{align}
We can then realize that $A_1$, $A_2$ and $A_3$ can be associated to $B$, $B'$ and $B''$ evaluated at $z\!=\!0$, or equivalently at $y\!=\!\frac{m}{v}$, in the following way
\begin{equation*}
\begin{array}{l}
B(x/\sqrt{v})=A'(0)=	A_1\ ,		\\
B'(x/\sqrt{v})=A''(0)= 2A_2\ ,		\\
B''(x/\sqrt{v})=A'''(0)=	6A_3\ ,
\end{array}	\qquad\qquad 	\text{with }x\equiv\frac{m}{\sqrt{v}}\ .
\end{equation*}
Thus we can establish the expression for $f$ in terms of $B$ and its derivatives,
\begin{equation}
f(k^2) = \frac{\pi_{1}-v\, A_{0}}{\pi_{0} -mA_0} = 
	-\frac vm +\frac{B''(x/\sqrt{v})-vx^2 B(x/\sqrt{v})}{B'(x/\sqrt{v})}
	\ .	\label{fdiBvm}
\end{equation}
We can then express the correlator in terms of Tricomi functions
\begin{equation}
\langle \ImO_\phi \ImO_\phi\rangle = i\;\frac{x\,(k^2-v)\left( 4v\, \sfU\Big[\frac{k^2+3v}{4v},\,\frac 12;\, x^{2} \Big] +(k^2+3v)\,\sfU\Big[\frac{k^2+7v}{4v},\,\frac 32;\, x^{2} \Big] \right)}{2\sqrt{v}\left( 2v\,\sfU\Big[\frac{k^2-v}{4v},\,\text{--}\frac 12;\, x^{2} \Big]  + (k^2-v)\,\sfU\Big[\frac{k^2+3v}{4v},\,\frac 12;\, x^{2} \Big] \right)}	\ .	\label{TriCorr}
\end{equation}

\begin{figure}[!htb]
	\centering
	\includegraphics[width=0.75\textwidth]{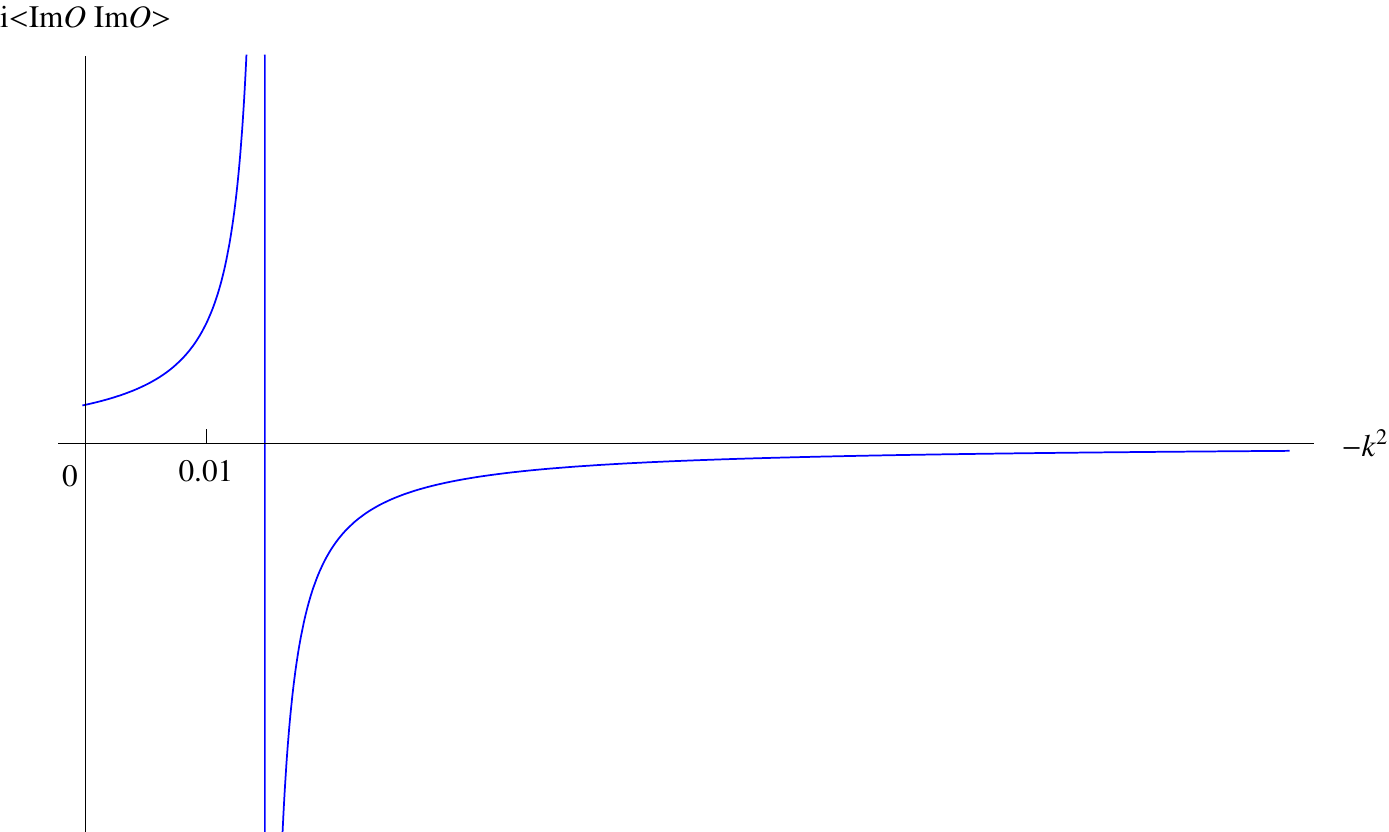}
	\caption{The lightest pole (PGB) in $\langle \ImO_\phi \ImO_\phi\rangle$, for $v\!=\!1$ and $x\!=\!0.01$, which is of the order of $x$.} \label{plotPGB1}
	\includegraphics[width=0.75\textwidth]{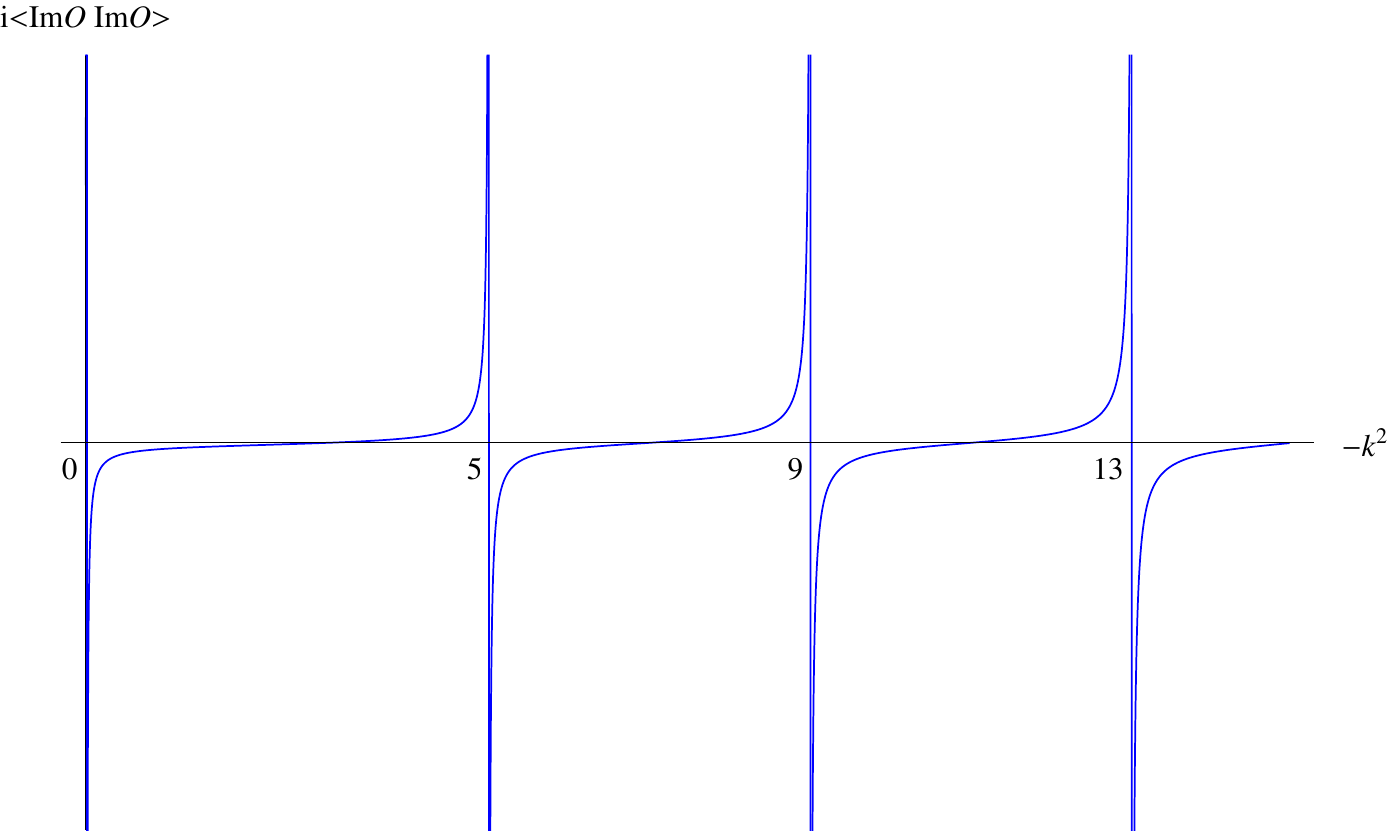}
	\caption{The first poles of the spectrum in $\langle \ImO_\phi \ImO_\phi\rangle$, for $v\!=\!1$ and $x\!=\!0.01$; they exhibit a gap of the order of $5v$ with respect to the first pole (PGB).} \label{plotPGB2}
\end{figure}

Let us show now how this expression reproduces all the physical features required by the field theory analysis.
First of all, in the limit of zero momenta, $f(k^2)$ as given in \eqref{fdiBvm} satisfies relation~\eqref{scheme}, i.e.~$f(0)=\frac vm$. This can be easily seen by using \eqref{ODEy} in order to obtain
\begin{equation}
f(k^2) = \frac vm +k^2 \frac{ B(x/\sqrt{v})}{B'(x/\sqrt{v})} \ .
\end{equation}
Moreover, we can graphically find the poles of the propagator by plotting the correlator for specific values of the ratio $x=\frac{m}{\sqrt{v}}$. 
For instance, with $x=0.01$, that is spontaneous breaking dominating on explicit breaking, we find a first pole close to zero (see fig.~\ref{plotPGB1}), 
and then a complete spectrum of higher massive poles with a gap considerably bigger than the mass of the first pole (see fig.~\ref{plotPGB2}). This is the hallmark of a pseudo-Goldstone boson. 
Furthermore, the gapped spectrum presents an interesting feature that we will show analytically for the purely spontaneous case in the next section: the poles are separated
by a regular gap in squared mass (except for the first higher pole after the PGB, which exhibits a slightly bigger gap from the rest of the spectrum). 
This is reminiscent of linear confinement.\footnote{ 
Indeed a phenomenological model like \cite{Karch:2006pv}, that is designed in order to achieve linear confinement, also ends up having confluent hypergeometric equations for the bulk fluctuations.}

In addition, we are able to analytically find the  GMOR linear relation~\eqref{gmor}. 
Indeed, one finds that the numerator of expression~\eqref{TriCorr} is just a constant in the limits $k^2\!\rightarrow\!0$ and $x\!\rightarrow\!0$ (taken in this order). 
If one then takes the denominator and expands it to the first order in $x$ and afterwards to the first order in $k^2$, it vanishes for
\begin{equation}\label{GMORpole}
k^2 = -2\sqrt{v}\, \frac{\Gamma[\frac 54]}{\Gamma[\frac 34]}\; m 	\ ,
\end{equation} 
where $\Gamma$ is the Euler function. So we have found the explicit value for the residue $\mu$ appearing in \eqref{gmor} for the specific model at hand, 
namely 
\begin{equation}
 \mu = \frac{v M^2}{m} = 2\, v^{3/2}\, \frac{\Gamma[\frac{5}{4}]}{\Gamma[\frac{3}{4}]}\ .
\end{equation}
We are also able to find  the deviations from the linear GMOR behavior to the desired order in $\frac{m}{\sqrt{v}}$, as we show in appendix~\ref{asfaras}.

\begin{figure}[!htb]
	\centering
	\includegraphics[width=0.75\textwidth]{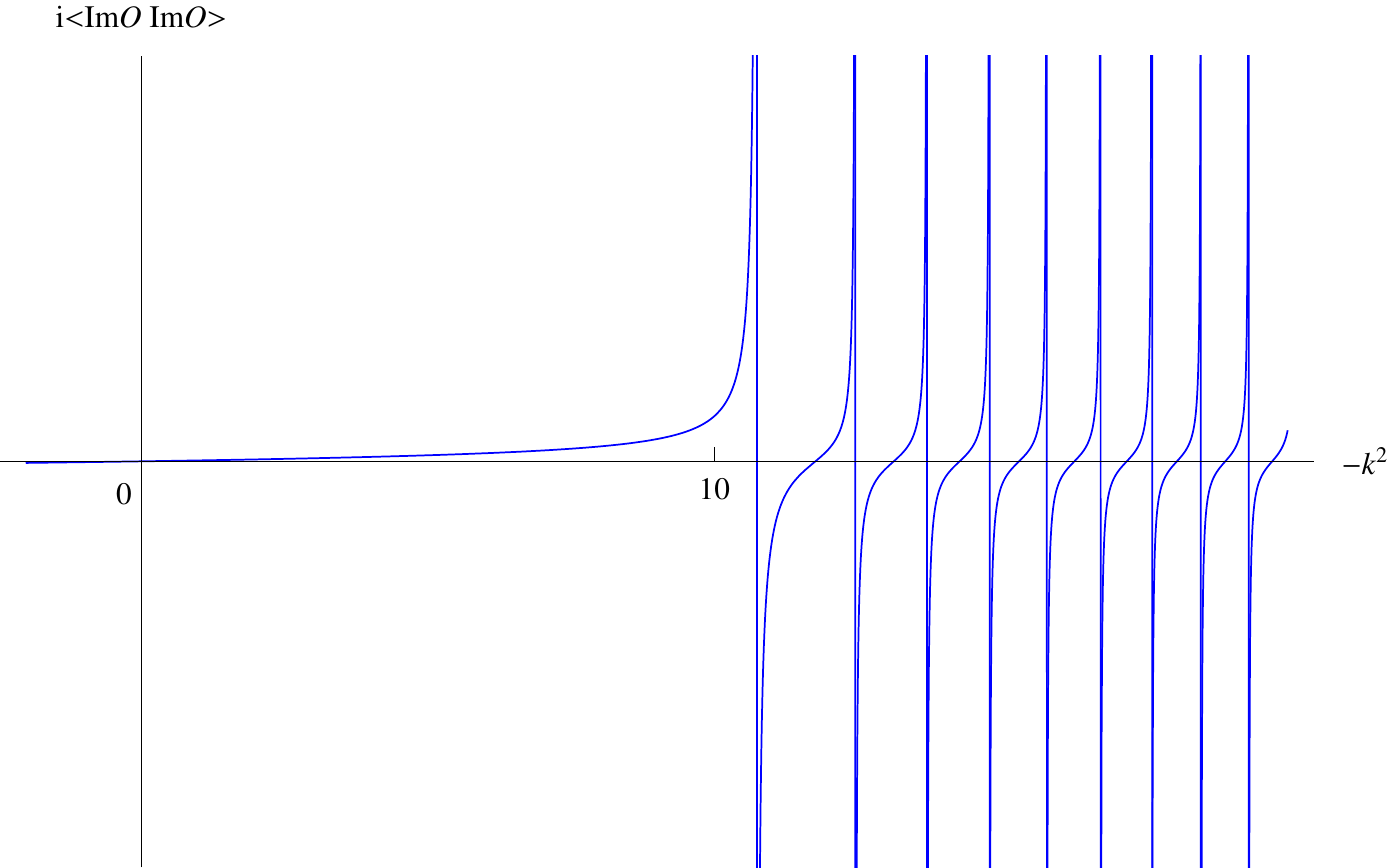}
	\caption{The low $|k^2|$ portion of the spectrum of $\langle \ImO_\phi \ImO_\phi\rangle$, for $v\!=\!0.1$ and $x\!=\!10$; the first pole is of the order of $m^2=100v=10$.} \label{plotPGBbigm}
\end{figure}

Let us underscore that expression~\eqref{fdiBvm} is valid not only for small $m$. We can then take $x\gg1$ and  find that, 
as expected, the first pole gets larger and larger with $m$ and it is pushed towards the rest of the spectrum, as can be seen in fig.~\ref{plotPGBbigm}. 
Actually, the ratio between the first gap and the subsequent ones increases with $x$, so that if one keeps the first pole fixed, the other 
poles will be increasingly dense just after it. This is the signal that a cut is emerging in the $x\to \infty$ limit, i.e.~in the purely explicit case.

In the next subsections we make further comments on the sub-cases of purely spontaneous or purely explicit breaking.

\subsection{Purely spontaneous case}		\label{purespont}

For purely spontaneous breaking, i.e.~$m=0$,  equation~\eqref{ODEnb} becomes
\begin{equation}
B''-\frac{2}{z}\,B'-k^2 B -v^2 z^2 B =0 \ .	\label{ODEvonly}
\end{equation}
This is the same as \eqref{ODEy}, but 
directly in the $z$ variable. Its solution is given in~\eqref{Bdiy}, where now $B$ is a function of $z$
\begin{equation}
B(z) =  C_1\: e^{-\frac{v}{2}z^2}\; \sfU\Big[\frac{k^2-v}{4v},\,-\frac 12;\, v z^{2} \Big]\ . 	\label{Bdiz}
\end{equation}

In this case $\pi_0$ is gauge invariant by itself, so
\begin{equation}
f(k^2)= \frac{\pi_1-vA_0}{\pi_0} \ ,
\end{equation}
and by using the equations of motion \eqref{eom-a} (for $F=1$ and $\phi_b=vz^2$), we obtain
\begin{equation}
\langle \ImO_\phi \ImO_\phi\rangle = -i\,f(k^2) = -i\,\frac{B'''(0)}{2B''(0)}=
	\frac{8i\,v^{\frac 32}}{k^2}\: \frac{\Gamma\Big[\frac{k^2+5v}{4v}\Big]}{\Gamma\Big[\frac{k^2-v}{4v}\Big]} 	\ ,	\label{fdiBv}
\end{equation}
which, in the limit of small momenta, actually exhibits a massless pole, signature of the expected Goldstone boson,
\begin{equation*}
\langle \ImO_\phi \ImO_\phi\rangle \sim - 2i\,v^{\frac 32}\, \frac{\Gamma[\frac 54]}{\Gamma[\frac 34]}\; \frac{1}{k^2}\ .
\end{equation*}
As a double check, we can recover the same result of eq.~\eqref{fdiBv} by taking the limit $m\!\rightarrow\!0$ of expression~\eqref{TriCorr}.
Furthermore, we are able to find explicitly the position of the gapped poles of the spectrum. Indeed, Gamma functions have no zeros, and they have poles at non-positive integer numbers. 
Therefore from~\eqref{fdiBv} we infer the following spectrum
\begin{equation*}
m_n^2=\left(5+4n\right) v \ ,
\end{equation*}
with $n$ being a non-negative integer. As anticipated in the previous section (see fig.~\ref{plotPGB2}), this spectrum presents the feature of equally gapped poles, 
except for the first massive one, whose gap from zero is bigger than the others by one unit in $v$.

\subsection{Purely explicit case}

For $v=0$ the equation reduces to
\begin{equation}
B''-k^2B-m^2B=0\ .
\end{equation}
Note that the limit of vanishing scalar profile is trivially achieved putting $m=0$ in the above equation, and in its solutions. 
The solutions are
\begin{equation}
B=C_+ e^{\sqrt{k^2+m^2}z}+C_- e^{-\sqrt{k^2+m^2}z}\ .
\end{equation}
Bulk normalizability imposes $C_+=0$. The gauge invariant combination is
\begin{equation}
\pi-mz A_l=-\frac{z}{m}\,B'=C_-\frac{\sqrt{k^2+ m^2}}{m}\left(z-z^2\,\sqrt{k^2+m^2}+\dots\right)\ .
\end{equation}
From this we read 
\begin{equation}
\pi_0-m A_0 = C_-\frac{\sqrt{k^2+m^2}}{m}\ , \qquad 
\pi_1 - m A_1 = -  C_-\frac{k^2+m^2}{m}\ .
\end{equation}
We can extract $A_1$ directly as the constant term of $B$, so that $A_1=C_-$. This gives
\begin{equation}
\pi_1 = - C_- \frac{k^2}{m} = - \frac{k^2}{\sqrt{k^2+m^2}} \left(\pi_0-m A_0\right)\ .
\end{equation}
Finally the correlator is given by
\begin{equation}
\langle \ImO_\phi \ImO_\phi\rangle = -i\,\frac{\delta\pi_1}{\delta\pi_0}=\frac{i\,k^2}{\sqrt{k^2+m^2}}\ .\label{opiopi}
\end{equation}
It presents a cut starting after a gap given by $m^2$. This is what was expected from the $m/\sqrt{v}\to \infty$ limit of the correlator in the general case. Note that when $m=0$ we obtain the conformal result $\langle \ImO_\phi \ImO_\phi\rangle\!=ik$, with a cut as well, but without any gap.

It is important also to notice that \eqref{opiopi} goes as $k^2$ for small $k$, which is necessary to ensure that the correlator $\langle J^\mu \ImO_\phi\rangle$ does not have a spurious massless pole.

\section{Conclusion}

The present paper systematizes knowledge that was in part already present in the literature but in a 
scattered fashion.  Our considerations  naturally split in three blocks. First we rely on purely quantum field theoretical 
arguments to determine the Ward identity structure expected on general grounds in the presence of a $U(1)$ symmetry breaking.
The analysis encompasses the generic case where the breaking can be explicit, spontaneous or concomitantly explicit and spontaneous.
Consistency arguments pinpoint the Ward identity structure independently of the strength of the coupling,  encoding 
the symmetry breaking pattern at the operatorial level. In particular, neither the explicit knowledge of the QFT Lagrangian nor that of the 
actual microscopic degrees of freedom are needed. This approach is able to encompass the generically composite nature 
of the fields responsible and emerging from the symmetry breaking at strong coupling.
Furthermore it allows for both a qualitative and quantitative control on the Goldstone modes and their pseudo relatives.
In fact, their masses and residues are constrained by the Ward identities and we show the validity in full generality of 
Gell-Mann-Oakes-Renner type of relations which relate the (pseudo)-Goldstone pole structure to the parameters of the symmetry breaking.

We then shift to holography to show how the Ward identity structure and symmetry breaking pattern can be 
neatly embodied in a simple and paradigmatic toy-model. The precise relations  among the correlators are realized by the holographic renormalization of the gauge/gravity model and rely on just an asymptotic near-boundary analysis.
This means that, in order to describe the Ward identities, only UV knowledge is necessary. The analysis can therefore be performed
before actually solving the model and discussing its IR properties. 

In turn, in order to access  quantitative data such as masses and residues, the IR properties are crucial and hence solving for bulk fluctuations becomes necessary.
We thus explicitly study the toy-model which allows for complete analytic control of its solutions and the dual correlators. 
Holography, already in one of its simplest realizations, is therefore able to reproduce general quantum field theoretical expectations
and allows for explicit quantitative computations. 
We expect that
the results of this analytic study remain qualitatively true in general for the entire class of toy-models 
in different space-time dimensions and featuring $U(1)$ breaking by means of charged operators of different scaling dimension.

The present study of pseudo-Goldstone modes has potentially far reaching future perspectives
when applied to other kinds of symmetries.
For instance, it would be interesting to consider supersymmetry breaking in holography along the lines of \cite{Argurio:2013uba,Argurio:2014uca,Bertolini:2015hua}, and the emergence of a putative pseudo-Goldstino. One should also  
consider non-relativistic set-ups, such as in the presence of temperature and/or chemical potentials. There is also the appealing possibility of considering directly the breaking of space-time symmetries which commute with the Hamiltonian. Regarding this latter possibility,
it is very interesting to study translation symmetry breaking.%
\footnote{Of particular interest in this context is the line of holographic studies initiated in \cite{Vegh:2013sk}
featuring specific mass terms for the bulk graviton which break spatial diffeomorphisms. Later similar models were
realized by means of neutral scalars through a Stueckelberg mechanism which allows for a Ward identity precisely accounting for
the translation breaking \cite{Andrade:2013gsa}. Along these lines, different further analyses have 
tackled or commented the possibility of having phonons in holography, see for instance \cite{Blake:2013owa,Baggioli:2014roa,Alberte:2015isw}.} This corresponds, when explicit, to 
a dual quantum field theory dissipating momentum and, when spontaneous, to a dual model featuring 
genuine phonon modes (roughly the Goldstone modes associated to translations). Of particular interest 
is the concomitant explicit and spontaneous case where momentum is dissipated and the system 
should possess pseudo-phonons. Its relevance is related to the effective description of condensed matter systems 
where heavy degrees of freedom (like impurities, disorder or lattices) are ubiquitous and essential 
to reproduce the correct phenomenology. At the same time,  one would like to have a clean theoretical control of
genuine phonons (or their pseudo-counterparts).

\section*{Acknowledgements}

We would like to thank Lasma Alberte, Andrea Amoretti, Daniel Arean, Matteo Baggioli, Matteo Bertolini, Francesco Bigazzi, Aldo Cotrone, Johanna Erdmenger, Manuela Kulaxizi, Nicodemo Magnoli, Andrei Parnachev, Oriol Pujolas, Diego Redigolo, Javier Tarrio
and Giovanni Villadoro for useful and enjoyable conversations and exchanges on the topics of the paper. RA and AMa would like to thank the Galileo Galilei Institute for hospitality during completion of this work. This research  is supported in part by IISN-Belgium (convention 4.4503.15), by the ``Communaut\'e
Fran\c{c}aise de Belgique" through the ARC program and by a ``Mandat d'Impulsion Scientifique" of the F.R.S.-FNRS. RA is a Senior Research Associate of the Fonds de la Recherche Scientifique--F.N.R.S. (Belgium).
The work of AMe ($\frac{\pounds}{2}$) is supported by the NWO Vidi grant.

\appendix 

\section{Generic $d$ and generic $\Delta$}

In this appendix we consider the analogues of eqs.~(\ref{eom-a})--(\ref{eom-pi}) in the case of generic $d$ and generic $\Delta$, given by $m^2_\phi=\Delta(\Delta-d)$.
The analysis shows that when we move away from the $d=3$, $\Delta=2$ case studied in the main text, an analytic treatment of the equations of motion, whenever possible, is more involved.
The generalized equations of motion read
%
%
\begin{align}
&
\frac{z^{d-1}}{F^{d/2-1}} 
\partial_z 
\left( 
z^{3-d} F^{d/2-1} \partial_z A_l	
\right)
+ 
\phi_b \pi
-
\phi_b^2 A_l 
= 0
\ ,\label{appAeq1}\\
&
\frac{z^2}{F} \Box \partial_z A_l 
-
\phi_b \partial_z \pi 
+ 
\partial_z \phi_b \pi = 0
\ ,\label{appAeq2}\\
&
\partial_z 
\left[ 
	\frac{F^{d/2}}{z^{d-1}} \partial_z \pi	
\right]
-
\frac{F^{d/2}}{z^{d+1}} m_\phi^2 \pi
+
\frac{F^{d/2-1}}{z^{d-1}} 
\left( 
	\Box \pi
	-
	\phi_b \Box A_l
\right)
= 0
\ . \label{appAeq3}
\end{align}
By defining 
\begin{align}
B \equiv z^{3-d} F^{d/2-1} \partial_z A_l \ ,
\label{appAdefB}
\end{align}
we can derive $\pi$ from eq.~(\ref{appAeq1}) and plug it into eq.~(\ref{appAeq2}), obtaining 
\begin{align}
z^2 B'' 
+
z \left[ 
	-2 z \frac{\phi_b'}{\phi_b} 
	+
	d - 1
	+
	\left( 1 - \frac{d}{2} \right) \frac{F'}{F} z
\right] B'
+
\left(
	z^2 \frac{\Box}{F} 
	-
	\phi_b^2 
\right) B
= 0
\ .
\label{appAeqB}
\end{align}
Therefore, by setting $F=1$ and by writing explicitly the scalar background
\begin{align}
\phi_b 
=
m z^{d-\Delta} + v z^\Delta
\ ,
\label{appAPhi}
\end{align}
we have
\begin{align}
&
z^2 \phi_b B''
+
 z
\left[ 
	m (2\Delta-d-1) z^{d-\Delta}
	+
	v (d-1-2\Delta) z^\Delta
\right] B'
+
\left( 
	 z^2\phi_b\Box
	-
	\phi_b^3 
 \right) B
= 0
\ .
\label{appAeqB2}
\end{align}
It is easy to see that the manipulations that have been used to recast \eqref{ODEnb} into a confluent hypergeometric form depend very much on $d=3$ and $\Delta=2$. 
Thus the generic case, namely in the presence of concomitant explicit and spontaneous breaking, will not have simple analytic solutions such as the ones of our toy-model. 

\section{Higher order corrections to GMOR} 	\label{asfaras}
In this appendix we discuss the corrections to the GMOR relation. First, 
we derive the  corrections in the small $m$ expansion of the GMOR relation given in (\ref{GMORpole}).
Using an iterative procedure, it is possible to obtain the analytic form of the first pole of the correlator (\ref{TriCorr}) up to the desired order in the small $m$ expansion.
To do so, we expand $k^2$ as follows
\begin{figure}[t]
	\centering
	\includegraphics[width=.75\textwidth]{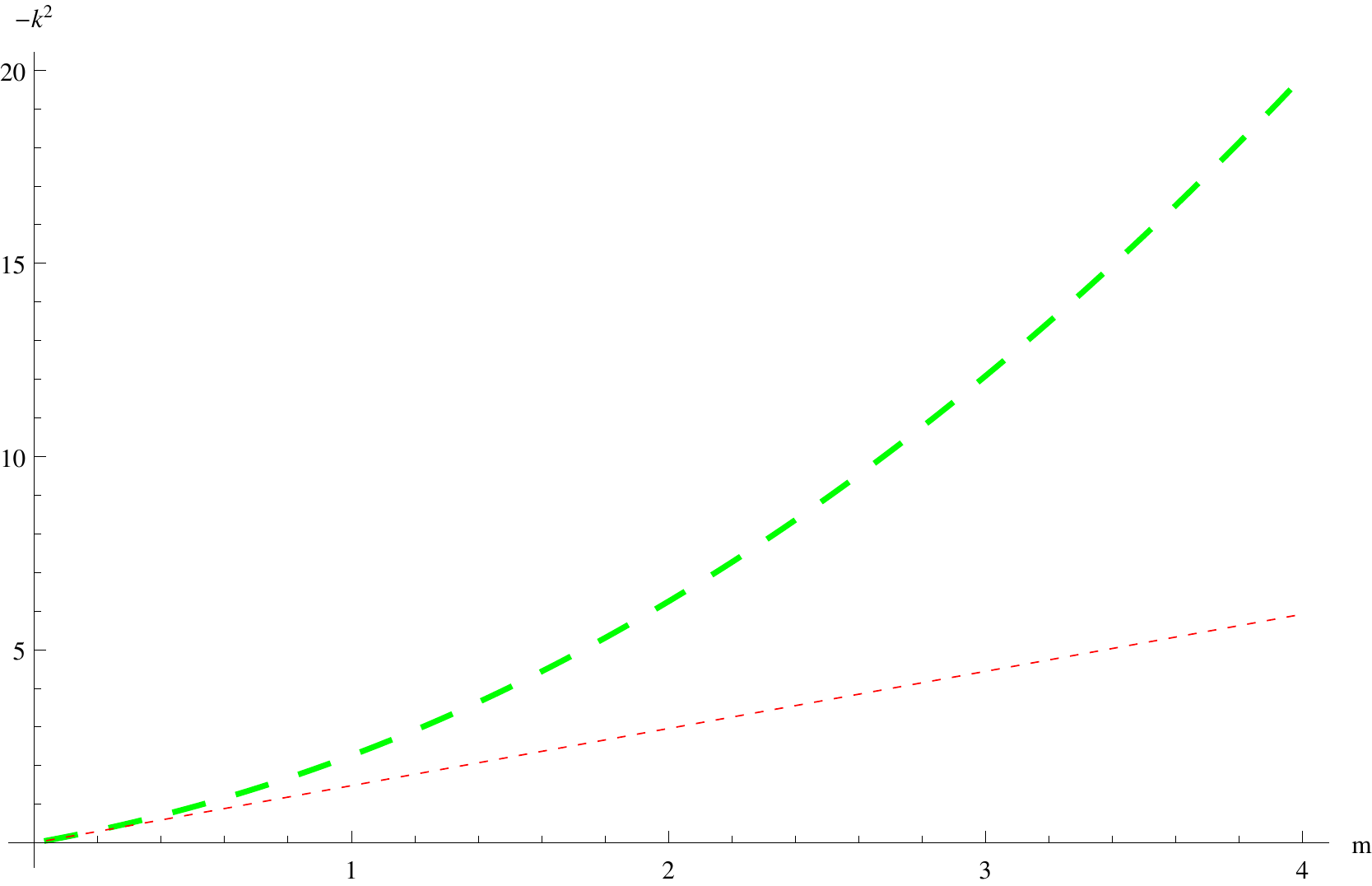}
	\caption{The short--dashed red curve represents the original GMOR relation (\ref{GMORpole}) and the large--dashed green line is the numerical value of the pole.
	%
	The plot is obtained by considering $v=1$.
	} \label{BppPlot}
\end{figure}
\begin{align}
k^2 
=
\sum_{i=1}^{N} \Xi_i \, m^i \ ,
\label{BppPol}
\end{align}
and we solve order by order the equation
\begin{align}
\langle \ImO_\phi \ImO_\phi\rangle^{-1} = 0 \ .
\end{align}
At first order we obtain the GMOR relation (\ref{GMORpole}) as expected, while by pushing further the analysis we obtain
\begin{align}
\Xi_2 
& =
1 - 
\frac{
	\pi \Gamma\left[ \frac 54 \right]^2
}{
\Gamma\left[ \frac 34 \right]^2
}
\ , \nonumber \\
\Xi_3
& =
2
\frac{
	\sqrt{2} \pi^2 \Gamma\left[ - \frac 14 \right]
	+
	8
	\left( 
	- 32
	+ 16 \, \mathfrak{c} 
	+ 3 \pi^2
	\right)
	\Gamma\left[ \frac 54 \right]^3
}{
\sqrt{v} \Gamma\left[ - \frac 14 \right]^3
} 
\ , \\
\Xi_4
& =
16 \pi^2
\frac{
	\left( 
		- 56
		+ 24 \, \mathfrak{c} 
		+ 3 \pi^2
	\right) \Gamma\left[ - \frac 14 \right]^2
	-
	16 \pi \left( 
		- 96
		+ 48 \, \mathfrak{c}
		+ 5 \pi^2
	\right) \Gamma\left[ \frac 54 \right]^2
}{
	3 v \Gamma\left[ - \frac 14 \right]^6
}
\ , \nonumber
\label{corrections}
\end{align}
where $\mathfrak{c}\simeq 0.915966$ is  Catalan's constant.

In turn, 
in fig.~\ref{BppPlot} we show how the numerical result for the first pole deviates from the original GMOR relation beyond the small $m/\sqrt v$ regime.

\end{document}